# *AB INITIO* STUDY OF LATTICE DYNAMICS OF DODECABORIDES $ZrB_{12}$ and $LuB_{12}$

N. M. Chtchelkatchev[a,b,*], M. V. Magnitskaya[a,c], E. S. Clementyev[d], and P. A. Alekseev[e]

[a] *Vereshchagin Institute for High Pressure Physics, Russian Academy of Sciences, Troitsk, Moscow, 108840 Russia*

[b] *Moscow Institute of Physics and Technology, Dolgoprudny, Moscow Region, 141700 Russia*

[c] *Lebedev Physical Institute, Russian Academy of Sciences, Moscow, 119991 Russia*

[d] *I. Kant Baltic Federal University, Kaliningrad, 236016 Russia*

[e] *National Research Center 'Kurchatov Institute', Moscow, 123182 Russia*

*e-mail: n.chtchelkatchev@gmail.com

**Abstract** — We report on *ab initio* study of lattice dynamics of frame-cluster dodecaborides $ZrB_{12}$ and $LuB_{12}$. Our calculated phonon frequencies and density of phonon states are consistent with available experimental and theoretical data.

## INTRODUCTION

Frame-cluster systems are complex compounds with a wide range of ground state types, which predetermines the functional properties and possibilities for considering them as promising materials. The lattice dynamics of such systems is not always amenable to calculations in the framework of widely known phenomenological models, which is established when comparing such model calculations with available experimental data. The reasons for this are obviously related to the nontrivial physics of interatomic interactions and electronic subsystem of these compounds.

One of the typical examples of such systems are dodecaborides of transition and rare-earth metals (TM and RE), where, along with the unusual physical properties of the d- and f-electron subsystem, a number of anomalies and features of lattice dynamics [1] are observed in the experimental phonon spectra. Obviously, the origin for this is the strong hierarchy of interatomic interactions and a large difference in the masses of constituents. On this basis, a rigid cluster frame is formed, which is stabilized by electrons entering the conduction band from the f (or d) electron shells of a heavy metal ion.

The state-of-the-art *ab initio* calculations of the phonon spectrum in TM and RE dodecaborides do not always describe the density of phonon states measured in neutron experiments well enough. Certain difficulties arise when trying to describe from first principles acoustic branches associated with vibrations of not only metal, but also boron ions, as well as the parameters of the energy gap between acoustic and optical parts and peaks of the phonon density of states near this gap (low-energy optical modes). The alternative simple phenomenological approach (based on effective force constants) also does not allow obtaining a consistent description of the results in the frequency range from 0 to 5–7 THz in the case of substitution in the metal sublattice. However, it is precisely this region that is of interest due to the unusual physical properties of these systems. The choice of appropriate numerical technique and the calculation of lattice dynamics in frame-cluster borides $ZrB_{12}$ and $LuB_{12}$ is the subject of this work.

## METHODS

Our *ab initio* computations are based on the density functional theory (DFT). We used the Quantum Espresso package with PAW-type pseudopotentials [2] and the PBE-GGA version of the exchange-correlation functional. We chose the plane-wave kinetic cut-off energy of 600 eV and uniform k-point grids for sampling the Brillouin zone with a reciprocal-space resolution of 0.08 Å$^{-1}$. The total energy convergence was better than $10^{-6}$ eV/cell. Phonon dispersions were computed using the density functional perturbation theory (DFPT) [2], with the interatomic force constants based on a 4×4×4 **q**-point grid.

## RESULTS AND DISCUSSION

Our calculated phonon dispersions and atom-projected phonon densities of states (PDOS) for $ZrB_{12}$ and $LuB_{12}$ are presented in Figs. 1 and 2, respectively. The main features of the phonon spectra are very similar for the two compounds. Generally, such shape of the phonon spectra is typical of rare-earth dodecaborides and has been repeatedly observed in neutron experiments for this class of compounds. Longitudinal and transverse acoustic modes are almost degenerate and characterized by low dispersion over large regions of **q**-space. The energy of these flat parts is about 3.8 and 3.3 THz for $ZrB_{12}$ and $LuB_{12}$, respectively. These parts correspond to a strong peak in the PDOS. In case of $ZrB_{12}$, the first peak is lower in magnitude (H = 5.36 $THz^{-1}$) and wider (HWHH = 0.34 THz) as compared to $LuB_{12}$ (5.58 $THz^{-1}$, 0.24 THz). The next peak is rather weak and situated at about 7 and 5.5 THz for $ZrB_{12}$ and $LuB_{12}$, respectively.

The first and second peaks are separated by a pronounced frequency 'gap' between the contributions of boron and metal atoms: the optical part of the phonon spectrum is determined mainly by B atoms, while acoustic modes are due to oscillations of heavier metal atoms. This picture corresponds to the frame-cluster representation of the vibrational spectra in these systems, which suggests that the vibrations of light and heavy atoms are separated. Our experiment-independent *ab initio* calculations explicitly show that there is an overlap of boron and metal vibrations. As is seen in Figs. 1 and 2, boron oscillations (pink curve) can contribute to the upper part of the acoustic modes, while oscillations of metal atoms (blue curve) can contribute to the lower part of the optical modes. Thus, at a strict quantitative level, we theoretically demonstrate the effect of mixing the eigenvectors of boron and metal vibrations – a conclusion that was previously made on the basis of experimental observations. In general, the width of the overlap region and the amplitude of the 'interference' contribution depend on the type of metal atoms. One can see that the contribution of metal atoms to the optical part of the spectrum and B atoms to the acoustic part is larger in case of $ZrB_{12}$.

The evaluated phonon dispersions and PDOS of $ZrB_{12}$ are in good agreement with the results of *ab initio* calculation [3]. It is worthwhile that a softening of the T1 transverse acoustic mode around the point X (001) obtained in [3] is confirmed in our calculation (see Fig. 1). Our theoretical phonon frequencies are also consistent with the experimental data of inelasting neutron scattering (INS) [3].

In case of $LuB_{12}$, however, the weaker second peak of PDOS related mainly to the boron vibrations and observed in the INS experiments [3] has not been obtained in the *ab initio* PDOS calculation [3]. In Fig. 2, we compare our theoretical results for $LuB_{12}$ with the INS data on PDOS [3] and phonon dispersions [4]. As is seen in the figure, there is a good agreement between the theoretical and experimental phonon frequencies and PDOS. Most likely, the second peak in the PDOS of LuB12 at 5.5 THz is absent in the *ab initio* study [3] due to insufficient numerical accuracy. The point is that the frozen-phonon approximation used in phonon calculations [3] requires rather large supercells, which is highly computationally demanding.

# CONCLUSIONS

We performed *ab initio* lattice-dynamics calculations of frame-cluster dodecaborides $ZrB_{12}$ and $LuB_{12}$. As a whole, our calculated phonon frequencies and atom-projected density of states are consistent with the results of available first-principles calculations and experimental measurements. So we conclude that the *ab initio* DFT approach is quite appropriate to study the sufficiently subtle physics of these compounds. Our experiment-independent calculations provide an explicit quantitative confirmation of mixing the eigenvectors of boron and metal vibrations, which was previously observed in experiments.

# ACKNOWLEDGMENTS

We thank D.A. Serebrennikov for technical assistance. This work was supported by Russian Science Foundation (Grant RSF 18-12-00438). The numerical calculations are carried out using computing resources of the federal collective usage center 'Complex for Simulation and Data Processing for Mega-science Facilities' at NRC 'Kurchatov Institute' (http://ckp.nrcki.ru/) and supercomputers at Joint Supercomputer Center of Russian Academy of Sciences (http://www.jscc.ru). We are also grateful for access to the URAN cluster (http://parallel.uran.ru) made by the Ural Branch of the Russian Academy of Sciences.

Figure captions.

**Fig. 1.** Calculated phonon dispersions (left) and atom-projected phonon density of states (right) for ZrB$_{12}$. Circles: experimental phonon frequencies from the INS data [3].

**Fig. 2.** Calculated phonon dispersions (left) and atom-projected phonon density of states (right) for LuB$_{12}$. Symbols: experimental phonon frequencies [4] and PDOS [3].

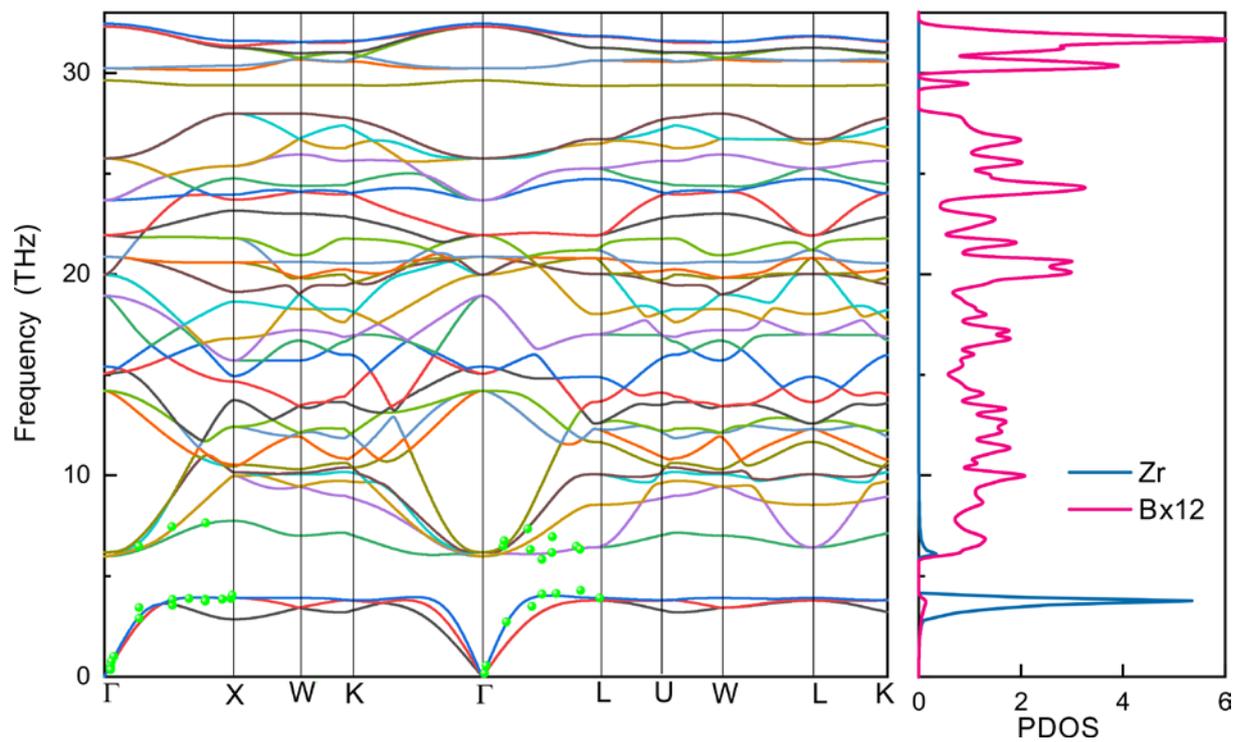

Fig. 1. N.M. Chtchelkatchev, *Journal of Surface Investigation*

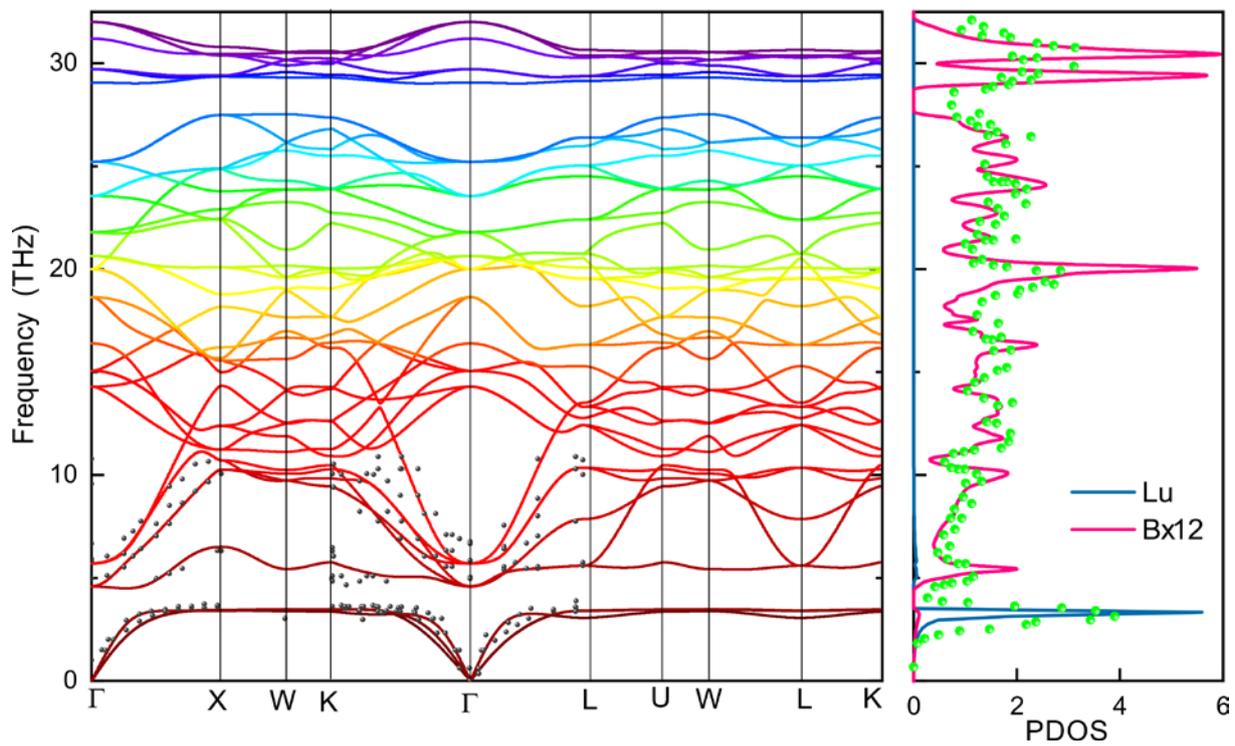

Fig. 2. N.M. Chtchelkatchev, *Journal of Surface Investigation*